\documentclass[10pt,conference]{IEEEtran}  

\IEEEoverridecommandlockouts       

\usepackage{graphicx}
\usepackage{cite} 
\usepackage{amsmath,amssymb,amsfonts,bm}
\usepackage[utf8]{inputenc}
\usepackage[T1]{fontenc}
\usepackage{xcolor}
\usepackage{algorithm}
\usepackage{algpseudocode}
\usepackage{subfig} 
\usepackage[belowskip=-5pt,aboveskip=0pt]{caption} 
\newcommand{\BS}{\textrm{BS}}
\newcommand{\RIS}{\textrm{RIS}}
\newcommand{\AOA}{\textrm{AOA}}
\newcommand{\AOD}{\textrm{AOD}}

\title{A New Channel Subspace Characterization for Channel Estimation in RIS-Aided Communications}

\author{Mehdi Haghshenas\textsuperscript{1},
Parisa Ramezani\textsuperscript{2}, Maurizio Magarini\textsuperscript{1}, Emil Bj{\"o}rnson\textsuperscript{2}\\
\IEEEauthorblockA{\textit{\textsuperscript{1}Department of Electronics, Information and Bioengineering, Politecnico di Milano, 20133 Milan, Italy}}
\IEEEauthorblockA{\textit{\textsuperscript{2}Department of Computer Science, KTH Royal Institute of Technology, SE-100 44 Stockholm, Sweden}\\ Email: \{mehdi.haghshenas, maurizio.magarini\}@polimi.it, \{parram,emilbjo\}@kth.se}\thanks{This work was supported by the FFL18-0277 grant from the Swedish Foundation for Strategic Research.}}

\begin{document}

\maketitle
\thispagestyle{empty}
\pagestyle{empty}

\begin{abstract}
A reconfigurable intelligent surface (RIS) is a holographic MIMO surface composed of a large number of passive elements that can induce adjustable phase shifts to the impinging waves. By creating virtual line-of-sight (LOS) paths between the transmitter and the receiver, RIS can be a game changer for millimeter-wave (mmWave) communication systems that typically suffer from severe signal attenuation. Reaping the benefits of RIS, however, relies on the accuracy of the channel estimation, which is a challenging task due to the large number of RIS elements. Specifically, conventional channel estimators require a pilot overhead equal to the number of RIS elements, which is impractical. Herein, we propose a novel way to approximately represent the RIS channels in a lower-dimensional subspace and derive the basis vectors for the identified subspace. We use this channel structure to only send pilots in this subspace, thereby vastly saving on the pilot overhead. Numerical results demonstrate that when the RIS has an element spacing of a quarter of the wavelength, our method reduces the pilot overhead by $80 \%$ with retained or even improved performance.

\end{abstract}
\begin{IEEEkeywords}
Holographic MIMO, reconfigurable intelligent surface, channel estimation, channel subspace characterization.
\end{IEEEkeywords}


\section{Introduction}
\label{sec:Intro}

Holographic MIMO (HMIMO) surfaces are software-controlled meta surfaces made of sub-wavelength elements that can collectively control the electromagnetic (EM) response of the surface and manipulate the attributes of the EM waves \cite{Huang2020}. Thanks to the recent advancements in micro electromechanical systems, the elements of HMIMO surfaces can be reconfigured dynamically and in real-time, thus catering to immediate changes of wireless networks. Among different types of these surfaces, the passive HMIMO surface, also known as reconfigurable intelligent surface (RIS), has recently attracted significant interest in academia and industry due to its low-cost and low-complexity design. Being able to co-phase the multi-path signals and reflect them towards the desired direction, RIS has been extensively studied for combating the blockage problem in wireless networks \cite{RuiToturial2021}. 

Millimeter wave (mmWave) systems are a specific example where the communication is impaired by blockage, resulting in a low signal-to-noise ratio (SNR) and unreliable connections \cite{ChenmmWave2018}. RIS can therefore play an important role in enhancing the performance of mmWave communication systems by providing strong virtual line-of-sight (LoS) paths between the transmitter and the receiver. 

To attain the promised functionality, accurate channel state information (CSI) of the cascaded channel from the transmitter to the receiver via the RIS is essential. Since the cascaded channel's pathloss is high per element, an RIS must compensate by having a massive number of elements and capitalizing on the quadratic beamforming gain \cite{RuiToturial2021}. Hence, the main challenge is how to efficiently estimate a large number of channel coefficients. Compressed sensing and least square (LS) are two widely-adopted signal processing techniques for channel estimation in RIS-aided systems \cite{ZhengSurvey2022}. Specifically, several works exploit the spatial structure and sparsity of the channel in the mmWave band to reduce the pilot overhead \cite{Li2020,ADMM2021,Haardt2021}. They convert channel estimation into sparse signal recovery problems and employ compressed sensing methods to solve them. These methods have high computational complexity for practically large RIS sizes and require strong assumptions on the angular separability between paths that might not be satisfied in practice. Moreover, compressed methods require high SNRs, which might not be available in mmWave bands and with the high pathloss of the cascaded channels. The conventional LS estimator is also impractical due to its prohibitive pilot overhead as it requires one pilot per RIS element \cite{Carvalho2020}. Another channel estimation approach has been recently investigated in \cite{Ozlem2022} that takes advantage of the existing spatial channel correlation in RIS-aided communications to reduce the pilot overhead.
Specifically, the authors evaluate the correlation matrix of the BS-RIS-user cascaded channel and identify a subspace of reduced dimension where all the cascaded channels approximately lie. 

In this paper, by exploiting the fact that the locations of the BS and the RIS are fixed, we propose a new method to further reduce the pilot length compared to \cite{Ozlem2022} that does not require the knowledge of spatial correlation matrix of the channel. We consider an RIS-aided mmWave communication system where the direct path between the base station (BS) and user is blocked. Since the BS and RIS are in fixed positions, the corresponding channel component is known in advance. We show that for an RIS with $M$ elements, the RIS-related channels approximately reside in a subspace of dimension $\eta$ with $\eta$$\,\ll\,$$M$. We first derive an orthogonal basis for this subspace and then develop a channel estimation framework where pilots are only transmitted in the derived subspace, while still enabling estimation of the entire cascaded channel and efficient RIS configuration for SNR maximization.
Numerical results show that, despite its lower pilot overhead and complexity, our method either outperforms LS or attains comparable performance depending on the transmit power. 

\section{System Model}
\label{sec:SysMod}

We consider a single-cell scenario where the BS serves single-antenna users through virtual LOS paths provided by the RIS. Accordingly, the RIS is deployed to have LOS paths toward the BS and all prospective user locations. The BS is equipped with a uniform linear array (ULA) of $N$ antennas, while the RIS is configured as a UPA with a total of $M$ antennas. The reflecting elements are arranged along the $yz$ axis with $M_\textrm{H}$ and $M_\textrm{V}$ elements on the $y$ and $z$ axes, respectively, such that $M = M_\textrm{H} M_\textrm{V}$. Since users are randomly distributed in the coverage area, their channels are unknown to the BS. Therefore, when a user wishes to connect to the BS, channel estimation should be performed at the BS prior to the data transmission. In the channel estimation phase, the UE transmits a pilot signal $x_t \in \mathbb{C}$ at time instance $t$. Assuming no direct path between the UE and BS, the received signal $\mathbf{y}_t \in \mathbb{C}^N$ at the BS can be modeled as \cite{Renzo2022s}
\begin{equation}
\label{eq:pmodelt}
    \mathbf{y}_t =  \mathbf{H} \mathrm{diag}(\bm{\phi}_t) \mathbf{h} x_t + \mathbf{n}_t = \mathbf{V} \bm{\phi}_t x_t + \mathbf{n}_t,
\end{equation}
where $\mathbf{n}_t \sim \mathcal{N}_{\mathbb{C}}(\mathbf{0},\sigma^2\mathbf{I}_N)$ is the additive independent complex Gaussian noise and the diagonal matrix $\mathrm{diag}(\bm{\phi}_t)$ contains the RIS phase shift configuration $\bm{\phi}_t = [e^{j\phi_{1,t}}, \dots, e^{j\phi_{M,t}}]^T$ used at time $t$. Additionally, $\mathbf{H} \in \mathbb{C}^{N\times M}$ is the channel between the BS and RIS, while $\mathbf{h} \in \mathbb{C}^{M}$ is channel between the RIS and UE. We define the cascaded BS-RIS-UE channel as $\mathbf{V} = \mathbf{H} \mathrm{diag}(\mathbf{h}) \in \mathbb{C}^{N \times M}$. 
We consider a mmWave band so the channels are modeled according to the geometric Saleh-Valenzuela channel model \cite{Molisch2014,Saleh1987}:
\begin{equation}
\label{eq:BSRISChan}
    \mathbf{H} = \sum_{\ell=1}^L\beta^{\ell} \mathbf{a}_{\BS}(\varphi_{\BS}^{\ell})\mathbf{a}_{\RIS}(\varphi_{\AOD}^{\ell},\theta_{\AOD}^{\ell})^H
\end{equation}
\begin{equation}
\label{eq:RISUEChan}
    \mathbf{h} = \sum_{\ell=1}^{L'}\zeta^{\ell} \mathbf{a}_{\RIS}(\varphi_{\AOA}^{\ell},\theta_{\AOA}^{\ell}),
\end{equation}
where the superscript $\ell$ corresponds to the $\ell$-th path in the multipath scenario, $L$ and $L'$ are the number of paths, $\beta$ and $\zeta$ are the complex channel gain coefficients, and $\mathbf{a}_{\BS}$ is the BS's array response vector defined as
\begin{equation}
    \begin{aligned}
        &\mathbf{a}_{\BS}(\varphi_{\BS}) = [1, e^{j2\pi d_{\BS} \sin(\varphi_{\BS})},\ldots, e^{j2\pi (N-1)d_{\BS} \sin(\varphi_{\BS})}]^{H},
    \end{aligned}
\end{equation}
where $\varphi_{\BS}$ is the azimuth-angle of arrival (AOA) of the RIS seen from the BS, $d_{\BS}$ is antenna spacing normalized by the carrier wavelength, and $\mathbf{a}_{\RIS}(\varphi,\theta)$ is the array response of the RIS defined as
\begin{equation}
\label{eq:UPARes}
    \begin{aligned}
        &\mathbf{a}_{\RIS} (\varphi,\theta) = \left [ 1,\ldots, e^{j2\pi[i(m) d_\textrm{H}\cos (\theta) \sin (\varphi) + j(m) d_\textrm{V} \sin (\theta)]} \right. \\
        & ,\ldots, \left. e^{j2\pi[(M_\textrm{H} - 1)d_\textrm{H}\cos (\theta) \sin (\varphi)+ (M_\textrm{V} - 1) d_\textrm{V} \sin (\theta)]} \right ]^H,
    \end{aligned}
\end{equation}
where the azimuth angle $\varphi$ and elevation angle $\theta$ seen from the RIS can be the angle of departure (AOD) towards the BS in \eqref{eq:BSRISChan}, and AOAs from the user in \eqref{eq:RISUEChan}.
Moreover, $ i(m)= \mathrm{mod}(m-1,M_\textrm{H})$ and $j(m) =  \lfloor \frac{m-1}{M_\textrm{H}}  \rfloor$ are the horizontal and vertical indicies of the $m$-th RIS element, and $\mathrm{mod}$ is the modulo operator. Similarly, $d_\textrm{H}$ and $d_\textrm{V}$ are the horizontal and vertical element spacing normalized by the carrier wavelength.

We consider transmission of $P$ pilot symbols, where $P$ will be specified later. Assuming the channels to be fixed during the estimation phase, $\bm{\Phi} = [\bm{\phi}_1, \ldots, \bm{\phi}_P] \in \mathbb{C}^{M \times P}$ collects the $P$ RIS configurations. Accordingly, we can write the collective received pilot symbols $\mathbf{Y} = [\mathbf{y}_1,\dots,\mathbf{y}_P] \in \mathbb{C} ^{N \times P}$ as
\begin{equation}
\label{eq:end2end}
    \mathbf{Y} = \mathbf{V}\mathbf{\Phi} \mathbf{X} + \mathbf{N},
\end{equation}
where $\mathbf{X} = \mathrm{diag}(x_1,\dots,x_P)$ and $\mathbf{N} = [\mathbf{n}_1,\dots, \mathbf{n}_P] \in \mathbb{C} ^{N \times P}$ are the collective transmitted pilot symbols and noise matrix, respectively. 
\section{Spanned Channel Subspace by the RIS}
\label{sec:Basis}

The channel coefficients are similar for adjacent RIS elements, and determined by the number of clusters and their angular distribution over the propagation environment. Even in an isotropic scattering environment, the spatial correlation between two RIS elements is a sinc function of the inter-element distance divided by $\lambda/2$ \cite{EmilRayleigh2021}. Hence, for a typical RIS with a planar structure and sub-$\lambda/2$ element spacing, there will be substantial spatial correlation under isotropic scattering.
The spatial correlation is even stronger in the mmWave band, which is highly non-isotropic with one dominant LOS path and a few additional paths.
One important implication of spatial correlation is that any $M$-dimensional RIS channel $\mathbf{h}$ of the type in \eqref{eq:RISUEChan} belongs to a subspace with a dimension substantially lower than $M$ \cite{EmilRayleigh2021}.
In this section, we demonstrate how the basis vectors for this subspace can be generated for a UPA.  

The channel in \eqref{eq:RISUEChan} is a linear combination of array response vectors and we will identify a set of orthogonal array response vectors that can be used to represent any such channel. There are multiple ways this can be done, but we
assume $\mathbf{a}_{\RIS}(\pi/2,0)$ is one of the vectors and will identify the remaining ones.
Using the expression of the array response for UPAs in \eqref{eq:UPARes}, we can write the inner product of two array response vectors obtained with the angle pairs $(\varphi_1,\theta_1)$ and $(\varphi_2,\theta_2)$ as \cite[Ch.~7]{EmilMIMObook}
\begin{equation} \label{eq:inner-product-arbitrary-angles}
    \begin{aligned}
        &\left |  \mathbf{a}(\varphi_2,\theta_2)^H\mathbf{a}(\varphi_1,\theta_1) \right | 
        =  \left | \sum_{m=1}^{M} e^{j2\pi  \left ( d_\textrm{V} j(m) \Omega+d_\textrm{H}i(m)\Psi \right )} \right | \\
        &= \underbrace{\left |   \sum_{k=0}^{M_\textrm{V}-1} e^{j2\pi d_\textrm{V} k\Omega}\right |}_{S(\Omega)} \underbrace{\left |  \sum_{l=0}^{M_\textrm{H}-1} e^{j2 \pi d_\textrm{H} l \Psi}\right |}_{T(\Psi)},
    \end{aligned}
\end{equation}
where
\begin{equation}
\label{eq:omega}
    \Omega = \sin(\theta_2) - \sin(\theta_1),
\end{equation}
\begin{equation}
\label{eq:PSi}
    \Psi = \cos(\theta_2)\sin(\varphi_2) - \cos(\theta_1)\sin(\varphi_1).
\end{equation}
Using standard techniques \cite[Ch.~7]{EmilMIMObook}, we can express $S(\Omega)$ and $T(\Psi)$ as
\begin{equation}
\label{eq:corr1}
    S(\Omega) = \left | \frac{\sin(\pi M_\textrm{V} d_\textrm{V} \Omega)}{M_\textrm{V} \sin(\pi d_\textrm{V} \Omega)}  \right |,
\end{equation}
\begin{equation}
\label{eq:corr2}
    T(\Psi) = \left | \frac{\sin(\pi M_\textrm{H} d_\textrm{H} \Psi)}{M_\textrm{H} \sin (\pi d_\textrm{H} \Psi)} \right |.
\end{equation}
We want to identify a set of angle pairs that result in mutually orthogonal array responses. The inner product in \eqref{eq:inner-product-arbitrary-angles} 
is zero if 
either $S(\Omega)$ or $T(\Psi)$ is zero. From \eqref{eq:corr1} we notice that $S(\frac{k}{M_\textrm{V} d_\textrm{V}}) = 0$ for $ k = \pm1, \dots,\pm (M_\textrm{V}-1)$, and $S(\Omega)$ is periodic with period $\frac{1}{d_\textrm{V}}$. 
Since we assumed that $(\pi/2,0)$ is one angle pair, we consider $\theta_1 = 0$ as a starting point and identify other elevation angles by solving $S(\Omega) = 0$. Since the function is periodic, for $ k = \pm1, \dots,\pm\left \lfloor \frac{M_\textrm{V}}{2}\right \rfloor$ we have $S(\Omega) = 0$ if 
\begin{equation}
    \Omega = \sin(\theta_2) = \frac{k}{M_\textrm{V} d_\textrm{V}},
\end{equation}
and solving for $\theta_2$, we obtain a set of elevation angles that are mutually orthogonal. After collecting all the elevation angles from the previous step in the set $\Theta = \left  \{ \theta_1, \ldots, \theta_n \right \}$, for each $\theta \in \Theta$ we also need to find the azimuth angles that result in $T(\Psi) = 0$. This is required to preserve the orthogonality among all the array response vectors with the same elevation angle. From \eqref{eq:corr2}, we observe that $T(\frac{l}{M_\textrm{H} d_\textrm{H}}) = 0$ for $l = \pm 1, \dots, \pm (M_\textrm{H}-1) $, and it is periodic with period $\frac{1}{d_\textrm{H}}$. Similarly, we consider $\varphi_1 = \pi/2$ as the reference point for each elevation angle $\theta_i \in \Theta$ to identify other azimuth angles that satisfy $T(\Psi) = 0$. Therefore, for $l = \pm 1, \dots, \pm (M_\textrm{H}-1)$, we construct the set
\begin{equation}
\label{eq:azSet}
    \varPhi(\theta_i) = \\
    \left \{ \varphi_2: \Psi = \cos(\theta_i)\left ( \sin(\varphi_2) - 1 \right ) = \frac{l}{M_\textrm{H} d_\textrm{H}} \right \}
\end{equation}
to collect all the azimuth angles that correspond to orthogonal beams associated with elevation angle of $\theta_i$.

\begin{figure}
    \centering
    \includegraphics[width = 0.9\linewidth]{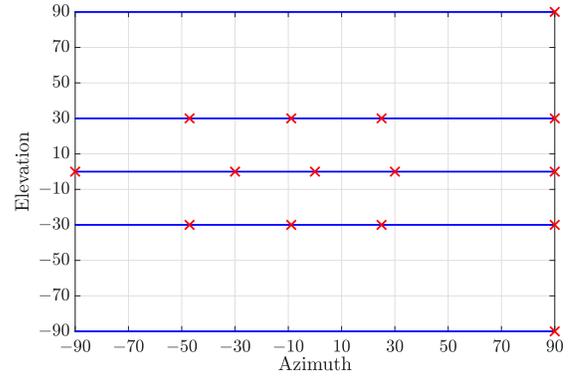}
    \caption{The points in the elevation-azimuth plane that give mutually orthogonal array response vectors when using an $8 \times 8$ RIS with $d_\textrm{H} = d_\textrm{V} = \frac{1}{4}$.}
    \label{fig:AzEl}
\end{figure}

In summary, we have determined a set of elevation angles that generates orthogonal channel directions. This step establishes the solid blue lines in the elevation-azimuth plane in Fig.~\ref{fig:AzEl} for an $8 \times 8$ RIS with $d_\textrm{H} = d_\textrm{V} = \frac{1}{4}$. In this figure, every two points belonging to different blue lines are mutually orthogonal. In the second step, for each blue line, we determine a collection of azimuth angles where any two choices satisfy the orthogonality condition \eqref{eq:azSet}. In Fig.~\ref{fig:AzEl}, these angles are shown with red crosses. We observe that for higher values of $|\theta|$, we have fewer points. This is because the beamwidth is inversely proportional to $\cos(\theta)$ and diverging from boresight where $\theta = 0$, the beamwidth increases \cite{balanis}. For any $(\varphi,\theta)$ in between the crosses in Fig.~\ref{fig:AzEl}, the corresponding array response vector can be expressed as a linear combination of the selected orthogonal array response vectors.

Following the above mentioned approach, the set $\mathcal{A} =\left  \{ (\varphi_1,\theta_1), \dots, (\varphi_{\eta},\theta_{\eta}) \right \}$ collects all the azimuth and elevation pairs (red crosses in Fig.~\ref{fig:AzEl}) that are generated in this way. 
Accordingly, set $ \mathcal{B} = \{ \mathbf{a}_{\RIS}(\varphi_{i},\theta_{i}): \hspace{5pt} (\varphi_i,\theta_i) \in \mathcal{A} \}$ collects all the orthogonal
array response vectors that form a basis of the subspace in $\mathbb{C}^M$ where the RIS channel resides. Algorithm~\ref{alg:Selection} summarizes the procedure to generate a set $\mathcal{B}$ of basis vectors.

\begin{figure}
    \centering
    \includegraphics[width = 0.9\linewidth]{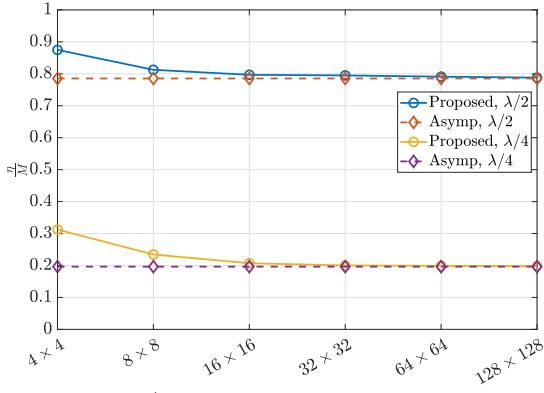}
    \caption{The ratio $\eta/M$ for different array sizes and element spacings.}
    \label{fig:Mratio}
\end{figure}

The cardinality $|\mathcal{A}| = |\mathcal{B}| = \eta$ defines the dimension of the subspace. This value is also known as the spatial degrees-of-freedom (DOF) and for a large planar aperture it can be asymptotically approximated as \cite{Sanguinetti2020, Edfors2018}
\begin{equation}
\label{eq:DOF}
     \eta \approx \pi M_\textrm{H} d_\textrm{H} M_\textrm{V} d_\textrm{V}.
\end{equation}
We notice that $\eta$ increases linearly with the area $M_\textrm{H} d_\textrm{H} M_\textrm{V} d_\textrm{V}$ of the aperture normalized by the wavelength.  To show under what conditions the approximate/asymptotic formula in \eqref{eq:DOF} 
is a good prediction of the number of basis vectors generated by Algorithm~\ref{alg:Selection}, Fig.~\ref{fig:Mratio} shows $\eta/M$ for the proposed algorithm and the approximate formula.
The dashed lines show the approximate DOF in \eqref{eq:DOF}, while the solid lines show the values obtained using the proposed algorithm. As it can be seen, the approximation in \eqref{eq:DOF} is tight for large arrays and the tightness appears earlier for arrays with small element spacings.

\begin{small}
\begin{algorithm}[!t]
\begin{algorithmic}[1]
\State Initialize $\mathcal{A} = \emptyset, \mathcal{B} = \emptyset$
\State $\Theta =  \{\theta = \arcsin(\frac{k}{M_\textrm{V} d_\textrm{V}})$
for $k = 0, \pm1, \ldots, \pm\left \lfloor \frac{M_\textrm{V}}{2} \right \rfloor \}$
\For{$\theta_i \in \Theta$}  
\For{$l = 0, \pm 1, \ldots, \pm (M_\textrm{H}-1)$}
    \State $\varphi \gets \arcsin\left ( 1 + \frac{\frac{l}{M_\textrm{H} d_\textrm{H}}}{\cos(\theta_i)} \right )$ 
    \State $\mathcal{A} \gets \mathcal{A} \cup \{ (\varphi,\theta_i)\}$
\EndFor
\EndFor
\For{($\varphi_i,\theta_i) \in \mathcal{A}$}
    \State $\mathcal{B} \gets \mathcal{B} \cup \mathbf{a}_{\RIS}(\varphi_i,\theta_i)$
\EndFor
\caption{The proposed algorithm to generate a set of basis vectors that spans the subspace of the RIS channel.}
\label{alg:Selection}
\end{algorithmic}
\end{algorithm}
\end{small}

\begin{figure}
    \centering
    \includegraphics[width =0.9 \linewidth]{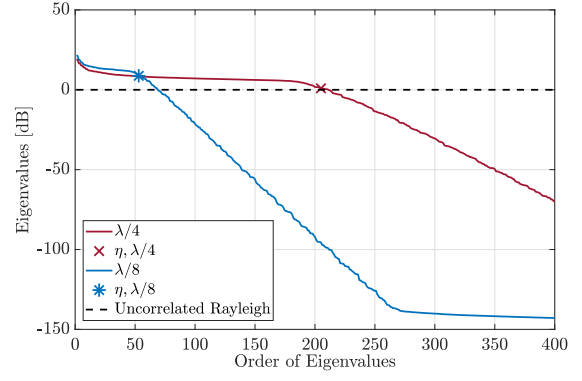}
    \caption{Eigenvalues in descending order for an RIS with $32 \times 32$ elements.}
    \label{fig:eig}
\end{figure}

Another observation from Fig.~\ref{fig:Mratio} is that $\eta < M$ and the ratio becomes particularly small when the element spacing is smaller than $\lambda/2$, which is the intended range for RIS.
The implication of this is that the channel $\mathbf{h} \in \mathbb{C}^M$ in \eqref{eq:RISUEChan} can be expressed as a linear combination of the $\eta$ orthogonal basis vectors $\mathbf{a}_{\RIS}(\varphi_i,\theta_i) \in \mathcal{B}$, such that
\begin{equation}
\label{eq:chanbasis}
    \mathbf{h} = \sum_{i=1}^{\eta} c_i \mathbf{a}_{\RIS}(\varphi_i,\theta_i),
\end{equation}
where $c_i$ is the channel coefficient associated with the basis vector $\mathbf{a}_{\RIS}(\varphi_i,\theta_i)$.
In Sec.~\ref{Sec:ChannelEst}, we will utilize this basis representation when performing channel estimation to reduce the required pilot length from $M$ to $\eta$. 
However, it should be noticed that \eqref{eq:chanbasis} is strictly speaking an approximate representation and there exists a tiny fraction of the signal power outside its span. To clarify further, we investigate the eigenvalues of the spatial channel correlation matrix $\mathbf{R} = \mathbb{E}\{ \mathbf{a}_{\RIS}(\varphi,\theta)\mathbf{a}_{\RIS}^H(\varphi,\theta) \}$ for a $32 \times 32$ RIS. Fig.~\ref{fig:eig} illustrates the eigenvalues of $\mathbf{R}$ in descending order where $\varphi \sim U[-\pi/3,\pi/3]$, $\theta \sim U[-\pi/2,\pi/2]$, and element spacings of $d_\textrm{H} = d_\textrm{V} \in \{\frac{1}{4},\frac{1}{8}\}$ are considered. The value of $\eta$ obtained by our algorithm is denoted by a star marker to specify the channel dimension that our proposed method considers. 
We observe that the number of large eigenvalues equals the number of basis functions in the representation \eqref{eq:chanbasis}, while the remaining eigenvalues decrease rapidly meaning that 98.7\% of the channel power lies in the identified $\eta$ dimensions. The remaining channel power can be neglected in practice, but cause a performance saturation at very large SNRs, which will be discussed further in Sec.~\ref{Sec:num}. 

\section{Channel Estimation and RIS Configuration}
\label{Sec:ChannelEst}
In this section, we first explain the intuition behind the RIS phase configuration and the  corresponding power gain at the BS considering perfect CSI. Then, we propose a novel method to perform channel estimation and RIS phase configuration by exploiting the channel parametrization in \eqref{eq:chanbasis}. 

\subsection{RIS Configuration with Perfect CSI and General Channels}

Suppose the signal $x$ is transmitted with power $\rho$. To recover it from the received signal vector $\mathbf{y}$ in \eqref{eq:pmodelt}, the BS applies the unit-norm receive combiner $\mathbf{w}$:
\begin{equation}
\label{eq:singlemod}
 y = \mathbf{w}^H\mathbf{H} \mathrm{diag}(\bm{\phi}) \mathbf{h} x + \mathbf{w}^H\mathbf{n} = \mathbf{w}^H\mathbf{V}\bm{\phi}x + \tilde{n},
\end{equation}
where $\mathbf{w}$ should be configured to maximize the SNR and $\tilde{n} = \mathbf{w}^H\mathbf{n} \sim \mathcal{N}_{\mathbb{C}}(0,\sigma^2)$. 
The achievable rate is $\log_2(1+\mathrm{SNR})$, where the SNR is 
\begin{equation}
    \mathrm{SNR} = \frac{ \rho |\mathbf{w}^H \mathbf{V} \bm{\phi} |^2}{\sigma^2}. \label{eq:SNR-general}
\end{equation}
The RIS phase configuration $\bm{\phi}$ and the receive combiner $\mathbf{w}$ should be jointly selected to maximize the SNR.
If the BS has perfect knowledge of the cascaded channel $\mathbf{V}$, it can compute the local optimal using the alternating optimization 
\begin{equation}
\label{eq:phiOpt}
\bm{\phi}_\mathrm{opt} = e^{j\cdot \arg(\mathbf{V}^H\mathbf{w})},
\end{equation} 
\begin{equation}
\label{eq:precOpt}
    \mathbf{w}_\mathrm{opt} = \frac{\mathbf{V}\bm{\phi}_\mathrm{opt}}{\|\mathbf{V}\bm{\phi}_\mathrm{opt} \|},
\end{equation}
and communicated to RIS via a backhaul connection \cite{RuiToturial2021}. 

Recall that the cascaded channel can be expressed as $\mathbf{V} = \mathbf{H} \mathrm{diag}(\mathbf{h})$. Since $\mathbf{w}^H \mathbf{V} \bm{\phi} =  \mathbf{w}^H\mathbf{H} \mathrm{diag}(\bm{\phi}) \mathbf{h}$, 
the optimized RIS configuration can be divided into two parts:
\begin{equation}
\label{eq:RISdiv}
 \bm{\phi} = \bm{\phi}_\mathrm{tx} \odot \bm{\phi}_\mathrm{rx},
\end{equation}
where $\odot$ denotes the Hadamard product. The part $\bm{\phi}_\mathrm{tx}$ acts as a transmit precoder towards the BS and $\bm{\phi}_\mathrm{rx}$ acts as receive combiner from the user. In the next subsection, we explain how these two parts should be designed to maximize the SNR in \eqref{eq:SNR-general} for LOS-dominant channels.

\subsection{RIS Configuration with Perfect CSI and LOS Channels}
\label{sec:Phaseintu}

We will now impose a practical structure on the channel. The BS and RIS are fixed in position and the channel is LOS-dominant in a mmWave band. Accordingly, we adopt the static LOS channel model to simplify the channel matrix in \eqref{eq:BSRISChan} to
\begin{equation}
\label{eq:BSRISChanv2}
    \mathbf{H} = \beta_{\mathrm{br}} \mathbf{a}_{\BS}(\varphi_{\BS})\mathbf{a}_{\RIS}(\varphi_{\AOD},\theta_{\AOD})^H,
\end{equation}
where $\beta_{\mathrm{br}}$ is the complex channel gain between RIS and BS.

Utilizing the fact that the channel matrix in \eqref{eq:BSRISChanv2} has rank one, we can express it as $\mathbf{H} = \mathbf{u}_1 \lambda_1 \mathbf{v}_1^H$, where $\mathbf{u}_1 = \frac{1}{\sqrt{N}} \mathbf{a}_{\BS}(\varphi_{\BS})$, $\lambda_1 = \sqrt{MN}\beta_{\mathrm{br}}$, and $\mathbf{v}_1 = \frac{1}{\sqrt{M}} \mathbf{a}_{\RIS}(\varphi_{\AOD},\theta_{\AOD})$. We can then rewrite \eqref{eq:singlemod} as
\begin{equation}
\label{eq:rxrank}
    y = \mathbf{w}^H \mathbf{u}_1 \lambda_1 \mathbf{v}_1^H \mathrm{diag}(\mathbf{h})\bm{\phi}x + \tilde{n} = \mathbf{w}^H \mathbf{u}_1 \tilde{\mathbf{v}}_1^H \bm{\phi} x+ \tilde{n},
\end{equation}
where $\tilde{\mathbf{v}}_1 = \lambda_1 \mathrm{diag}(\mathbf{h}^{*})\mathbf{v}_1$. Accordingly,
the SNR in \eqref{eq:SNR-general} can be simplified as
\begin{equation}
    \mathrm{SNR} = \frac{ \rho |\mathbf{w}^H \mathbf{u}_1 |^2  |\tilde{\mathbf{v}}_1^H \bm{\phi}  |^2}{\sigma^2}. \label{eq:SNR-simplfied}
\end{equation}
This expression enables separate optimization of the receive combining $\mathbf{w}$ and the RIS configuration $\bm{\phi}$. The maximum SNR is achieved when $\mathbf{w} = \mathbf{u}_1$ \cite{EmilMIMObook} and the unit-modulus entries of $\bm{\phi}$ have the same phase as the corresponding entries in $\tilde{\mathbf{v}}_1$.
The latter corresponds to maximizing $| \tilde{\mathbf{v}}_1^H\bm{\phi} |^2 = |\lambda_1|^2 | \mathbf{v}_1^H \mathrm{diag}(\mathbf{h}) \bm{\phi} |^2 = |\lambda_1|^2 | \mathbf{v}_1^H \mathrm{diag}(\bm{\phi}_\mathrm{tx}) \mathrm{diag}(\bm{\phi}_\mathrm{rx}) \mathbf{h} |^2$, which is achieved by $\bm{\phi}_\mathrm{tx} = \sqrt{M}\mathbf{v}_1$ and $\bm{\phi}_\mathrm{rx} = e^{-j \cdot \arg(\mathbf{h})}$.

In case of a LOS channel $\mathbf{h} = \zeta \mathbf{a}_{\RIS}(\varphi_{\AOA},\theta_{\AOA})$, then $|\mathbf{w}^H \mathbf{u}_1 |^2= 1$ and $| \tilde{\mathbf{v}}_1^H\bm{\phi} |^2 = M |\lambda_1 \zeta|^2 = M^2N|\beta_{\mathrm{br}} \zeta|^2$. Hence, under perfect CSI, the SNR is proportional to $N$ and to $M^2$. This is the behavior to strive for with imperfect CSI.

\subsection{Joint Channel Estimation and Phase Shift Optimization}
Upon the network deployment, the angles $\varphi_{\BS}, \varphi_{\AOD}$, and $\theta_{\AOD}$ in \eqref{eq:BSRISChanv2} are fixed and known to the BS. Following the discussion in Sec.~\ref{sec:Phaseintu}, the vectors $\mathbf{w}$ and $\bm{\phi}_\mathrm{tx}$ can then be selected optimally as $\mathbf{w} = \mathbf{u}_1 = \frac{1}{\sqrt{N}}\mathbf{a}_{\BS}(\varphi_{\BS})$ and $\bm{\phi}_\mathrm{tx} = \sqrt{M} \mathbf{v}_1 = \mathbf{a}_{\RIS}(\varphi_{\AOD},\theta_{\AOD})$. The key practical challenge is to select $\bm{\phi}_\mathrm{rx} = e^{-j \arg(\mathbf{h})}$ when the UE-RIS channel $\mathbf{h}$ is unknown a priori.
To estimate the unknown part of the cascaded channel, the UE transmits a constant pilot signal $x = \sqrt{\rho}$ with power $\rho$ at $\eta$ time instances. We configure the RIS phase shift vector at each time instance as
\begin{equation}
    \bm{\phi}_t = \bm{\phi}_\mathrm{tx} \odot \mathbf{a}_{\RIS}(\varphi_t,\theta_t),
\end{equation}
by considering the set of orthogonal basis vectors $\mathbf{a}_{\RIS}(\varphi_t,\theta_t) \in \mathcal{B}$ derived in Sec.~\ref{sec:Basis}.
By adding time indices on the received signal, RIS configuration, and noise, we can rewrite \eqref{eq:rxrank} for $t=1,\ldots,\eta$ as
\begin{equation}
\label{eq:1symbchan}
     y_t = \sqrt{\rho}\mathbf{w}^H \mathbf{u}_1 \tilde{\mathbf{v}}_1^H\bm{\phi}_t + \tilde{n}_t = \sqrt{\rho}\mathbf{w}^H \mathbf{V} \bm{\phi}_t + \tilde{n}_t,
\end{equation}
where the scalar $\mathbf{w}^H \mathbf{V} \bm{\phi}_t$ is the projection of $\mathbf{w}^H \mathbf{V}$ onto $\bm{\phi}_t$. Since  $\mathbf{w}=\mathbf{u}_1$, it follows that $\mathbf{w}^H\mathbf{u}_1 = 1$ and $y_t$ therefore corresponds to the projection of $\tilde{\mathbf{v}}_1$ on $\bm{\phi}_t$. 
By collecting the received signals in \eqref{eq:1symbchan} in the vector $\mathbf{y} \in \mathbb{C}^\eta$ and the RIS configurations in $\mathbf{\Phi} = [\bm{\phi}_1, \cdots, \bm{\phi}_\eta] \in \mathbb{C}^{M\times \eta}$, we obtain
\begin{equation}
\label{eq:accumu}
     \mathbf{y}^T = \sqrt{\rho} \tilde{\mathbf{v}}_1^H \mathbf{\Phi}  + \tilde{\mathbf{n}}^T.
\end{equation} 
We can compute the reduced-subspace LS (RS-LS) estimate of $\tilde{\mathbf{v}}_1$ in the subspace spanned by the columns of $\mathbf{\Phi}$ as
\begin{equation}
\label{eq:estimatedChan}
     \hat{\tilde{\mathbf{v}}}_1 = \frac{\mathbf{\Phi}\mathbf{y}^*}{M\sqrt{\rho}}  = \frac{1}{M} \mathbf{\Phi}\mathbf{\Phi}^H\tilde{\mathbf{v}}_1 + \frac{1}{M\sqrt{\rho}}\mathbf{\Phi} \tilde{\mathbf{n}}^* ,
 \end{equation}
where $\frac{1}{M}\mathbf{\Phi}\mathbf{\Phi}^H$ is the projection matrix onto the subspace of dimension $\eta$ that we derived in Sec.~\ref{sec:Basis}. Since we know that $\tilde{\mathbf{v}}_1$ approximately falls into this subspace, the projection matrix $\frac{1}{M}\mathbf{\Phi}\mathbf{\Phi}^H$ basically works as an identity matrix. However, the noise term $\frac{1}{M\sqrt{\rho}}\mathbf{\Phi} \tilde{\mathbf{n}}^*$ has the covariance matrix  $\frac{\sigma^2}{\rho M^2} \mathbf{\Phi} \mathbf{\Phi}^H$ with the trace $\frac{\sigma^2}{\rho } \frac{\eta}{M}$, thus all the noise outside the signal subspace is removed which improves the estimation quality.
The resulting estimate of the entire cascaded channel is
\begin{equation}
\label{eq:Vclosed}
    \hat{\mathbf{V}} = \mathbf{u}_1 \hat{\tilde{\mathbf{v}}}_1^H = \mathbf{w} \hat{\tilde{\mathbf{v}}}_1^H.
\end{equation} 
In line with Sec.~\ref{sec:Phaseintu}, we select the phase configuration as
\begin{equation}
\label{eq:phiclosed}
     \hat{\bm{\phi}} = e^{j \cdot \arg (\hat{\tilde{\mathbf{v}}}_1)}.
\end{equation}
\section{Numerical Evaluation}
\label{Sec:num}

We consider a LOS simulation scenario where the user can be anywhere in the RIS coverage area specified by $\varphi \in [-\pi/3,\pi/3]$, $\theta \in [-\pi /2 , 0]$, and  distance $d \in [30, 50] $ meters with uniform probability. The carrier frequency is  $28$ GHz, the noise power is $\sigma^2 = -96$ dBm, and the bandwidth is $20$ MHz. The BS is equipped with a ULA of $128$ antennas with antenna spacing $d_{\BS} = 1/2 $. The RIS is equipped with a UPA with $M_\mathrm{H} = M_\mathrm{V} =128$ elements, and the element spacing $d_\mathrm{H}=d_\mathrm{V}=\frac{1}{4}$. 
We will vary the transmit power $\rho$.

The LOS propagation channel between the RIS and BS is generated according to \eqref{eq:BSRISChanv2}, where $\beta_{\mathrm{br}} = \beta_0 e^{-j \frac{2 \pi}{\lambda}d_{\mathrm{br}}}$ 
with $\beta_0 = -81.4$ dB corresponding to the distance of $d_{\mathrm{br}} = 10$\,m \cite{Rappaport2014}. For simplicity we assume the RIS is deployed at the same elevation as the BS such that the BS see the RIS at the boresight; that is, $\varphi_{\BS} = \theta_{\AOD} = 0$ and $\varphi_{\AOD} =\pi/6$.

We adopt a correlated Rician fading model to generate the channel between the RIS and UE. The Rician factor (the ratio of the power of the LOS component to all the NLOS components) is evaluated based on $K = 13 - 0.03 \cdot \frac{d}{1\, \mathrm{m}}$  [dB], where $d$ is the distance between the RIS and UE \cite{3gpp.25.996}. Initially, the LOS complex coefficient is evaluated as $\beta_0  e^{-j \frac{2 \pi}{\lambda}d}$, where $\beta_0= -61.4 - 20\log_{10}(d / 1\, \mathrm{m})$ dB \cite{Rappaport2014}. Accordingly, for the NLOS paths $\ell \neq 1$, we generate the complex channel coefficient as
    $\zeta^{\ell} \sim \mathcal{N}_{\mathbb{C}}(0,\gamma_{\ell})$,
where the associated power $\gamma_{\ell}$ is evaluated based on the parameters reported in \cite{Rappaport2014}.

\begin{figure}
    \centering
    \includegraphics[width = 0.9\linewidth]{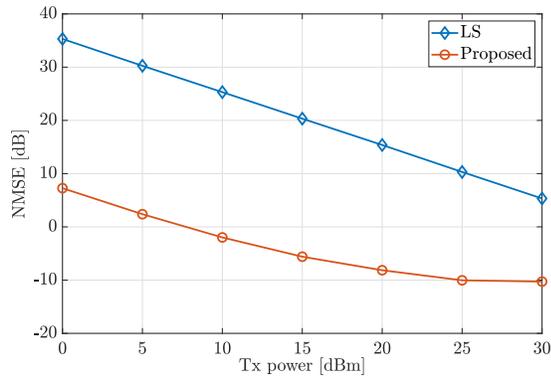}
    \caption{NMSE versus pilot transmit power.}
    \label{fig:NMSE128}
\end{figure}

We will now evaluate the performance of the proposed approach that we described in Sec.~\ref{Sec:ChannelEst}, where we estimated the channel using \eqref{eq:Vclosed} and configured $\bm{\phi}$ according to \eqref{eq:phiclosed}. We use the conventional LS estimator from \cite{Carvalho2020} as the benchmark. The receive combining and RIS configuration are then computed through the alternating optimization in \eqref{eq:phiOpt} and \eqref{eq:precOpt} by treating the LS estimate as being perfect. For an estimated cascaded channel $\hat{\mathbf{V}}$ we define the normalized mean-squared-error (NMSE) as
\begin{equation}
    \textrm{NMSE} = \frac{1}{K} \sum_{i=1}^{K} \frac{\| \mathbf{V}_i - \hat{\mathbf{V}}_i\|^2_{\mathrm{F}}}{\| \mathbf{V}_i\|^2_{\mathrm{F}}},
\end{equation}
where $K$ is the number of realizations in the Monte Carlo simulation, and $\| \cdot \|_{\mathrm{F}}$ is the Frobenius norm.
Fig.~\ref{fig:NMSE128} shows the  NMSE as a function of the pilot transmit power $\rho$, averaged over $K=500$ user locations. According to \eqref{eq:estimatedChan}, our designed estimator reduces the total noise power by a factor $\eta / M$ compared to the LS estimator, without reducing the total received signal power. In this setup, $\eta / M  \approx 0.198$. The figure shows that our proposed scheme vastly outperforms LS in terms of NMSE although the pilot length in our method is only $19.8$\% of that with the LS estimator.
Moreover, we notice that the NMSE with our estimator approaches an error floor at very high transmit power. We anticipated such behavior in Sec.~\ref{sec:Basis} since our proposed algorithm discards channel dimensions containing very low power and only accounts for the significant dimensions. Nevertheless, the attained accuracy is adequate to configure the RIS phase shift very well.

\begin{figure}
    \centering
    \includegraphics[width=0.9\linewidth]{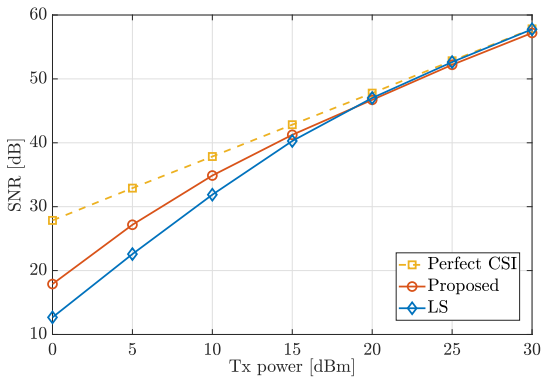}
    \caption{Average SNR with respect to the transmit power.}
    \label{fig:SNR128}
\end{figure}

Fig.~\ref{fig:SNR128} shows the average SNR for data transmission in \eqref{eq:SNR-general} with respect to the pilot transmit power.
We compare our proposed method with the maximum SNR obtained with perfect CSI  and with the LS estimator as before. By increasing the transmit power, the gap between the optimal and achieved SNR decreases. This is due to the higher SNR during the channel estimation. Our method outperforms LS when the power budget is limited while the LS estimator is slightly better at very high transmit power values. This is a consequence of the fact that a tiny fraction of the channel power falls outside the proposed subspace considered in the channel estimation. Such a loss is acceptable considering that our method reduced the pilot overhead by $80.2\%$ compared to the LS estimator.
This overhead saving improves the overall data rate as more time is dedicated to data transmission. 

\section{Conclusion}
\label{sec:Conc}

In this paper, we proposed a resource-efficient joint channel estimation and RIS phase configuration method that exploits the fact that the channels involving the RIS approximately reside in a low-dimensional subspace. We first characterized this subspace by deriving a set of basis vectors that spans it and used these bases to obtain a set of RIS configurations to consider during channel estimation.
We demonstrated that our method outperforms the conventional LS estimator in terms of channel estimation accuracy while requiring a shorter pilot length. For example, when the element spacing is a quarter of the wavelength, the pilot overhead can be reduced by $80 \%$.

\bibliography{Reference}
\addtolength{\textheight}{-12cm}   

\end{document}